\documentclass[prb,twocolumn,floatfix,amsmath,nofootinbib]{revtex4-1}

\usepackage{graphicx}
\usepackage{color}
\usepackage{dsfont}
\usepackage[normalem]{ulem}
\usepackage{dcolumn}
\usepackage{multirow}
\usepackage{textcomp}
\usepackage{placeins}
\usepackage{subfigure}
\usepackage{float}
\hfuzz=\maxdimen
\newcommand{\eg}{e.g.,\@ }
\newcommand{\ie}{i.e.\@ }
\newcommand{\cf}{cf.\@ }
\newcommand{\etal}{\textit{et al.\@ }}
\newcommand{\abinitio}{\textit{ab initio}}
\newcommand{\abohr}{a_\mathrm{B}}

\begin{document}

\title{First-principles investigation of chiral magnetic properties in multilayers: Rh/Co/Pt and Pd/Co/Pt}
\author{Hongying Jia}
\email[Corresponding author: ]{h.jia@fz-juelich.de} 
\author{Bernd Zimmermann}
\email[Corresponding author: ]{be.zimmermann@fz-juelich.de} 
\author{Stefan Bl\"ugel}
\affiliation{Peter Gr\"unberg Institut and Institute for Advanced Simulation, Forschungszentrum J\"ulich and JARA, 52425 J\"ulich, Germany}

\date{\today}

\begin{abstract}
The magnetic properties of (111) oriented Rh/Co/Pt and Pd/Co/Pt multilayers are investigated by first-principles calculations. We focus on the interlayer exchange coupling, and identify thicknesses and compositions where a typical ferromagnet or a synthetic antiferromagnet (SAF) across the spacer layer is formed. All systems under investigation show a collinear magnetic intralayer order, but the Dzyaloshinskii-Moriya interaction (DMI) is rather strong for Pd-based systems, so that single magnetic skyrmions can be expected. In general, we find a strong sensitivity of the magnetic parameters (especially the DMI) in Rh-based systems, but Pd-based multilayers are less sensitive to structural details.    
\end{abstract}

\maketitle
\section{Introduction}

The Dzyaloshinskii-Moriya interaction (DMI)\cite{1Dzialoshinskii,2Moriya,3Moriya} is an antisymmetric exchange interaction, which exists in magnetic systems that lack inversion symmetry and exhibit (strong) spin-orbit coupling (SOC). It may play an important role in determining the physical properties of surfaces and interfaces of low-dimensional metallic magnets\cite{4Bland,5Blugel,6Soumyanarayanan,7Bode,8Yang,9Heide,Romming}, in particular for the formation of chiral magnetization textures such as domain walls, spin spirals, and skyrmions\cite{10Wiesendanger,11Heinze,12Romming,13Chen,14Boulle,Perini,Grab}.

Although originally proposed in bulk materials\cite{1Dzialoshinskii}, the interface induced DMI in low-dimensional metallic magnets can be much stronger. Recently, in an one-dimensional monoatomic Mn chain deposited at the Pt(664) step-edge, a spiral magnetic ground state was conjectured to be induced by a large DMI\cite{15Schweflinghaus}, and a spiral magnetic ground state was observed in Fe chains on Ir(001) witnessing a DMI energy, which is even as large as the Heisenberg exchange interaction energy\cite{16Menzel}. Furthermore, small isolated skyrmions are found as metastable states at low temperatures in ultra-thin magnetic films, which are in contact with a non-magnetic metallic layer with a large SOC\cite{17Romming,18Sampaio,19Fert,Herve}. 

In order to stabilize skyrmions at room temperature, major attention has recently been focused on magnetic multilayers (MMLs), where a larger magnetic volume increases the thermal stability, and the repetitive interfaces allow for additive DMI facilitating the formation of chiral textures\cite{20Luchaire,21Dupe,22Soumyanarayanan,23Nandy,Maccariello,Yagil}. Additionally, MMLs provide the flexibility to design materials and tune the Dzyaloshinskii-Moriya interaction, exchange stiffness and magnetocrystalline anisotropy through the choice of different metals at the interfaces\cite{21Dupe,22Soumyanarayanan,Cao}. A very illustrative example was the theoretical investigation of the Fe based  \{$4d$/Fe/$5d$\} multilayers\cite{21Dupe}, structures in which Fe layers are sandwiched between $4d$ and $5d$ transition-metal layers.  It was noticed that in these structures the exchange and the Dzyaloshinskii-Moriya interactions that control the skyrmion formation as well as the size can be tuned separately by the two different interfaces with a $4d$ and $5d$ metal.

In this work, we explore the properties of the experimentally more vital Co/Pt-based magnetic multilayers. We selected Rh/Co/Pt and Pd/Co/Pt multilayers with Rh being isoelectronic to Co and Pd isoelectronic to Pt. We determine key magnetic interactions, such as interlayer exchange coupling (IEC), magnetocrystalline anisotropy, exchange stiffness and Dzyaloshinskii-Moriya interaction for various layer thicknesses by density-functional theory (DFT) calculations. We explore the possibility to tune these properties by varying the thickness of the Rh, Pd and Pt  between 1 and 5 layers. Since most multilayers are grown by sputtering techniques resulting in (111) textured growth with (111) oriented interfaces\cite{14Boulle,20Luchaire,Zeissler}, we choose fcc(111) oriented layers with C$_{3v}$ symmetry with the in-plane lattice constant fixed to the one of Pt ($a = 5.24~\abohr = 277~\mathrm{pm}$) and subsequently optimize the structure along the out-of-plane direction (\ie along the $z$-axis; see Sec.~\ref{sec:results:structure} for details).

We find that adding one more Pd or Pt layer to the smallest MML increases the perpendicular magnetic anisotropy (PMA) by more than 30\%, while adding a Co leads to a reduction of PMA. We find that the choice of the 4$d$ transition metal element has a significant effect on the sign of the IEC between Co layers. Furthermore, our results show that Pt atoms give the largest contributions to the total DMI, but its sign and magnitude is very sensitive with respect to the number of atomic layers as well as the choice of the $4d$ element (Rh or Pd).

\section{Magnetic model and parameters from DFT}

From a magnetic viewpoint, the multilayers under investigation are composed of individual magnetic layers $n$ which are separated by non-magnetic spacers. The magnetic layers interact with each other via dipolar fields and interlayer exchange interaction,
\begin{equation}
  E(\{m_n\}) = - \frac{1}{2} \sum_n J_{0n} ~ \hat{\mathbf{m}}_0 \cdot \hat{\mathbf{m}}_n \, , \label{eq:iec:ham}
\end{equation}
where $\hat{\mathbf{m}}_n$ refers to the magnetic moment of layer $n$ taken at unit length and $J_{0n}$ are respective exchange constants. The interlayer exchange-coupling energy,
\begin{equation}
  E_\mathrm{IEC} =  E_\mathrm{SAF} - E_\mathrm{FM} \, ,\label{eq:iec:energy}
\end{equation}
indicates whether the synthetic antiferromagnetic (SAF) or ferromagnetic (FM) state is lower in energy at zero magnetic field.

Each individual magnetic layer may be described by a continuous vector magnetization $\mathbf{m}_n$ in the framework of the micromagnetic model\cite{24Heide,25Aharoni}, where the energy reads 
\begin{equation}
E[\mathbf{m}_n] = \int \mathrm{d}^2 r \left[\frac{A}{4\pi^2} (\dot{\mathbf{m}}_n^2) + \frac{1}{2\pi}  \mathcal{D}:\mathcal{L}(\mathbf{m}_n) + \mathbf{m}_n^T \mathcal{K} \mathbf{m}_n\right],
\label{eq:1}
\end{equation}
where $A$ is the exchange stiffness, $\mathcal{D}$ is the spiralization tensor, $\mathcal{L}(
\mathbf{m})=\nabla \mathbf{m} \times \mathbf{m}$ is the chirality tensor and $\mathcal{K}$ is the magnetocrystalline anisotropy tensor. Here, the symbol $\mathcal{D}:\mathcal{L} = \sum_{\alpha,\beta} \mathcal{D}_{\alpha \beta} \, \mathcal{L}_{\alpha \beta}$ denotes a contraction of two tensors. For our case of (111) textured multilayers, the spiralization tensor $\mathcal{D}_{\alpha \beta}=D\,\varepsilon_{\alpha \beta}$ only depends on a single DMI parameter $D$ and effects only N\`eel-type spin-spirals. $\varepsilon_{\alpha \beta}$ represents the antisymmetric Levi-Civita tensor. The determination of the parameters $A$ and $D$ is based on the calculation of the energy of homogeneous spin spirals, for which Eq.~\ref{eq:1} simplifies to

\begin{equation}
  \frac{E(\lambda)}{\lambda} = \frac{A}{\lambda^2} + \frac{D}{\lambda} + \frac{K}{2} \,.
  \label{eq:2}
\end{equation}
The left-hand side can be conveniently calculated from DFT by evaluating the total energy for a set of spin-spirals with different period lengths $\lambda = 2\pi/\vert \mathbf{q} \vert$, where $\mathbf{q}$ is a spin-spiral propagation vector parallel to the film plane. $A$ and $D$ are then obtained from quadratic and linear fits in $\lambda^{-1}$.  Details about the calculation methods can be found in Refs.~\onlinecite{15Schweflinghaus,27Heide,28Zimmermann}.

It is convenient to introduce a reduced dimensionless parameter,
\begin{equation}
  \kappa = \left(\frac{4}{\pi}\right)^2 \frac{AK}{D^2} \, .
  \label{eq:3}
\end{equation}
If $\kappa \in [0,1)$, the magnetic structure in each layer exhibits a periodic spin spiral as a magnetic ground-state, with increasing inhomogeneity as $\kappa$ approaches $1$. For $\kappa > 1$, the layer will exhibit a collinear magnetic structure\cite{24Heide,29Dzyaloshinskii,30Izyumov}. This expression is particularly applicable for low temperatures and without external fields. While the spin-spiral state is a one-dimensional chiral magnetic structure, skyrmions are two-dimensional ones. Neglecting the 
stray field and rescaling length and energy scales, the energy functional Eq.~\ref{eq:1} can brought into a form where $\kappa$ enters as only parameter\cite{Lux}. Hence, the metastability and the profile of skyrmions are determined qualitatively by $\kappa$. First estimates indicate that metastability can be obtained of values of $\kappa$ much larger than unity.

\section{Methods and Computational details}

The DFT calculations have been performed using the full-potential linearized augmented plane-wave (FLAPW) method, as implemented in the FLEUR code\cite{31www}. The structural optimizations have been carried out applying the scalar-relativistic approximation with a mixed (LDA/GGA) exchange-correlation functional\cite{32Santis}: the local density approximation (LDA)\cite{33Vosko} was used in the muffin-tin (MT) spheres of Pt, whereas the generalized gradient approximation (GGA)\cite{34Perdew} was employed in the other regions, \ie in the interstitial region and MT spheres of Co, Rh, and Pd. The ferromagnetic order was assumed for structural relaxations. For the calculation of magnetic parameters the  LDA has been used. For all calculations, we chose the radii of MT spheres as $2.2~\abohr$ for Co and Rh, $2.3~\abohr$ for Pd and $2.5~\abohr$ for Pt, where $\abohr$ is the Bohr radius. The LAPW basis functions included all wave vectors up to $k_\mathrm{max} = 4.0~\abohr^{-1}$ in the interstitial region and in the MT spheres, basis functions including spherical harmonics up to $l_\mathrm{max} = 10$ were taken into account.

\subsection{Interlayer exchange coupling}

In order to determine the interlayer exchange coupling (IEC), we perform spin-spiral calculations in scalar relativistic approximation for spin-spiral vectors $\mathbf{q}$ along the high symmetry line $\Gamma$-A of the Brillouin zone (see Fig.~\ref{fig1}c), where $\Gamma$ represents the ferromagnetic and A the synthetic antiferromagnetic state. In order to get correct energies for states with $\mathbf{q} \neq 0$, it is important to relax the direction of induced magnetic moments during the self-consistent calculations. The full Brillouin zone was sampled by ($24\times24\times10$) $k$-points.

For the determination of the model parameters (see Eq.~\ref{eq:iec:ham}), we performed least squares fits including nearest and next-nearest neighbor exchange constants ($J_1$ and $J_2$). In this model, the interlayer exchange coupling energy (\cf Eq.~\ref{eq:iec:energy}) is identical to the nearest-neighbor exchange constant,
\begin{equation}
  E_\mathrm{IEC} = J_1 \, .
\end{equation}

\subsection{Magnetic anisotropy energy}

The magnetic anisotropy is composed of the magnetocrystalline anisotropy due to the spin-orbit interaction and the dipolar energy due to the classical magnetic dipole-dipole interactions. The dipolar energy is calculated straightforwardly assuming magnetic moments on a lattice with the dipolar energy summed up by an Ewald summation \cite{9Heide,Draaisma}.

In order to obtain the magnetocrystalline anisotropy energy (MCA), self-consistent relativistic calculations with SOC for magnetizations along the $z$ axis in the FM/SAF ground states were first performed. We converged the charge density until self-consistency was achieved using ($48\times48\times20$) $\mathbf{k}$-points to integrate the Brillouin zone (BZ). Regarding the difference in the magnetization directions as a perturbation, Andersen's force theorem (FT)\cite{35Mackintosh,36swald,37Liechtenstein} was employed to calculate the energy difference between the magnetization directions along the $z$ axis and the in-plane $x$ axis. The MCA can therefore be approximated by a summation over all occupied (occ.) states as
\begin{equation}
E_\mathrm{MCA} \approx \sum_{\mathbf{k}\nu}^{\mathrm{occ.}}{\varepsilon_{\mathbf{k}\nu}^{\mathrm{FT}}(\mathbf{\hat{e}}_x)} - \sum_{\mathbf{k}\nu}^{\mathrm{occ.}}{\varepsilon_{\mathbf{k}\nu}^{0}(\mathbf{\hat{e}}_z)},
\label{eq:4}
\end{equation}
where $\nu$ is the band index, $\mathbf{k}$ is the Bloch vector, $\mathbf{\hat{e}}$ denotes the magnetization direction, and $\varepsilon_{\mathbf{k}\nu}^{0}$ and $\varepsilon_{\mathbf{k}\nu}^{\mathrm{FT}}$ are the spectra of the unperturbed and perturbed Hamiltonians, respectively.

\subsection{Spin-stiffness}

The spin stiffness $A$ is dominated by non-relativistic interactions of electrons. In this case the spin-spiral is a stationary magnetic state, whose energy is calculated efficiently employing the  generalized Bloch theorem\cite{39Sandratskii}. Hence, the energy $E(\lambda)=E_\mathrm{SS}(\mathbf{q})$ of homogeneous spin spirals with wave vector $\mathbf{q}$ is calculated according to the following steps: First, we obtain a self-consistent charge density in scalar-relativistic approximation for the collinear ground state $\mathbf{q}_0$ (FM or SAF) using ($24\times24\times10$) $\mathbf{k}$-points in the full Brillouin zone. Second, we use this charge density to calculate spin-spiral energies for $\mathbf{q}$-vectors in the vicinity of the ground state employing the force theorem of Andersen\cite{35Mackintosh,36swald,37Liechtenstein}. The error of the spin-spiral energy estimated by comparison to self-consistent calculations is in the order of 5\%. Then, we extract $A$ by a quadratic fit of the spin-spiral energies $E_\mathrm{SS}(\mathbf{q}) \propto A\left| \mathbf{q}_\mathrm{eff} \right|^2$, where $\mathbf{q}_\mathrm{eff} = \mathbf{q} - \mathbf{q}_0$ is the change in the spin-spiral vector from the ground state. Calculations are performed for $\mathbf{q}_\mathrm{eff}$ covering 20\% of BZ of the spin-spiral wavevectors and using $48\times48\times20$ $\mathbf{k}$-points in the BZ of the Bloch states.

\subsection{Dzyaloshinskii-Moriya interaction}

The DMI arises from relativistic spin-orbit coupling in an inversion asymmetric crystal field. Due to the symmetry of the MMLs studied here ($C_{3v}$ symmetry), the DMI affects the energy of N\'{e}el-type magnetic structures (as opposed to Bloch-type). Hence, we calculate the SOC-induced change in energy of cycloidal spin-spirals. Since SOC effects are small as compared to the other contributions to the Hamiltonian\cite{26Lezaic}, we employ first-order perturbation theory to include SOC on-top of a scalar-relativistic spin-spiral calculation. The energy change reads
\begin{equation}
  E_\mathrm{DMI}(\mathbf{q}) = \sum_{\mathbf{k}\nu}{n_{\mathbf{k}\nu}(\mathbf{q})\,\delta\varepsilon_{\mathbf{k}\nu}(\mathbf{q})},
    \label{eq:5}
\end{equation}
where $\mathbf{k}$ is the Bloch vector, $\nu$ is the band index, $n_{\mathbf{k}\nu}(\mathbf{q})$ is the occupation number of the scalar-relativistic state $\vert \mathbf{k},\nu \rangle$, and $\delta\varepsilon_{\mathbf{k}\nu}(\mathbf{q}) = \langle \mathbf{k},\nu \vert \mathcal{H}_\mathrm{so}\vert \mathbf{k}, \nu \rangle$ is the spin-orbit induced shift of band-energy of this state in first order perturbation theory. The same $\mathbf{q}_\mathrm{eff}$ and $\mathbf{k}$-points as in the calculation of the spin stiffness are used. The values of $D$ are then extracted as the linear part of a cubic fit to the energy, \ie $E_\mathrm{DMI}(\mathbf{q}) = \frac{D}{2\pi} \vert\mathbf{q}_\mathrm{eff}\vert + C \vert\mathbf{q}_\mathrm{eff}\vert^3$. In the micromagnetic limit $\mathbf{q}_\mathrm{eff} \rightarrow 0$, the linear part will dominate.

\section{Results and Discussion}
\subsection{Structural Properties}\label{sec:results:structure}

\begin{table*}
 \caption{The stacking sequence, the equilibrium lattice parameter $c$ along $z$ axis, and the distances between different atomic layers $d$ ($d_\mathrm{double}$ represents the distance between atomic layers of the same chemical element in the unit cell). The number in brackets denotes the number of atomic layers, and $\abohr$ is the Bohr radius. \label{table1}}
 \begin{ruledtabular}
  \begin{tabular}{cccccccc}
       & Systems & \multicolumn{1}{c}{Stacking} & $c$ [$\abohr$] & \multicolumn{1}{c}{$d_{4d-\mathrm{Co}}$} & \multicolumn{1}{c}{$d_{\mathrm{Co-Pt}}$} & \multicolumn{1}{c}{$d_{\mathrm{Pt}-4d}$} & \multicolumn{1}{c}{$d_{\mathrm{double}}$} \\
       &         & \multicolumn{1}{c}{sequence} &                      & \multicolumn{1}{c}{[$\abohr$]}   & \multicolumn{1}{c}{[$\abohr$]}   & \multicolumn{1}{c}{[$\abohr$]}   & \multicolumn{1}{c}{[$\abohr$]}  \\ \hline 
  \multirow{4}{*}{Rh} & \{Rh(1)/Co(1)/Pt(1)\} & ABC  & 12.07 & 3.78 & 3.93 & 4.35 & ---  \\
                      & \{Rh(1)/Co(1)/Pt(2)\} & ABCB & 16.30 & 3.76 & 3.84 & 4.30 & 4.40 \\
                      & \{Rh(1)/Co(2)/Pt(1)\} & ABAB & 15.54 & 3.80 & 3.91 & 4.35 & 3.49 \\
                      & \{Rh(2)/Co(1)/Pt(1)\} & ABAC & 16.08 & 3.80 & 3.87 & 4.28 & 4.13 \\ \hline
  \multirow{4}{*}{Pd} & \{Pd(1)/Co(1)/Pt(1)\} & ABC  & 12.20 & 3.91 & 3.86 & 4.44 & ---  \\
                      & \{Pd(1)/Co(1)/Pt(2)\} & ABCB & 16.46 & 3.86 & 3.82 & 4.39 & 4.40 \\
                      & \{Pd(1)/Co(2)/Pt(1)\} & ABAB & 15.74 & 3.93 & 3.89 & 4.47 & 3.45 \\
                      & \{Pd(2)/Co(1)/Pt(1)\} & ABAC & 16.51 & 3.87 & 3.83 & 4.41 & 4.40 \\
  \end{tabular}
 \end{ruledtabular}
\end{table*}

\begin{figure}[H]
\centering
\includegraphics[width=70mm]{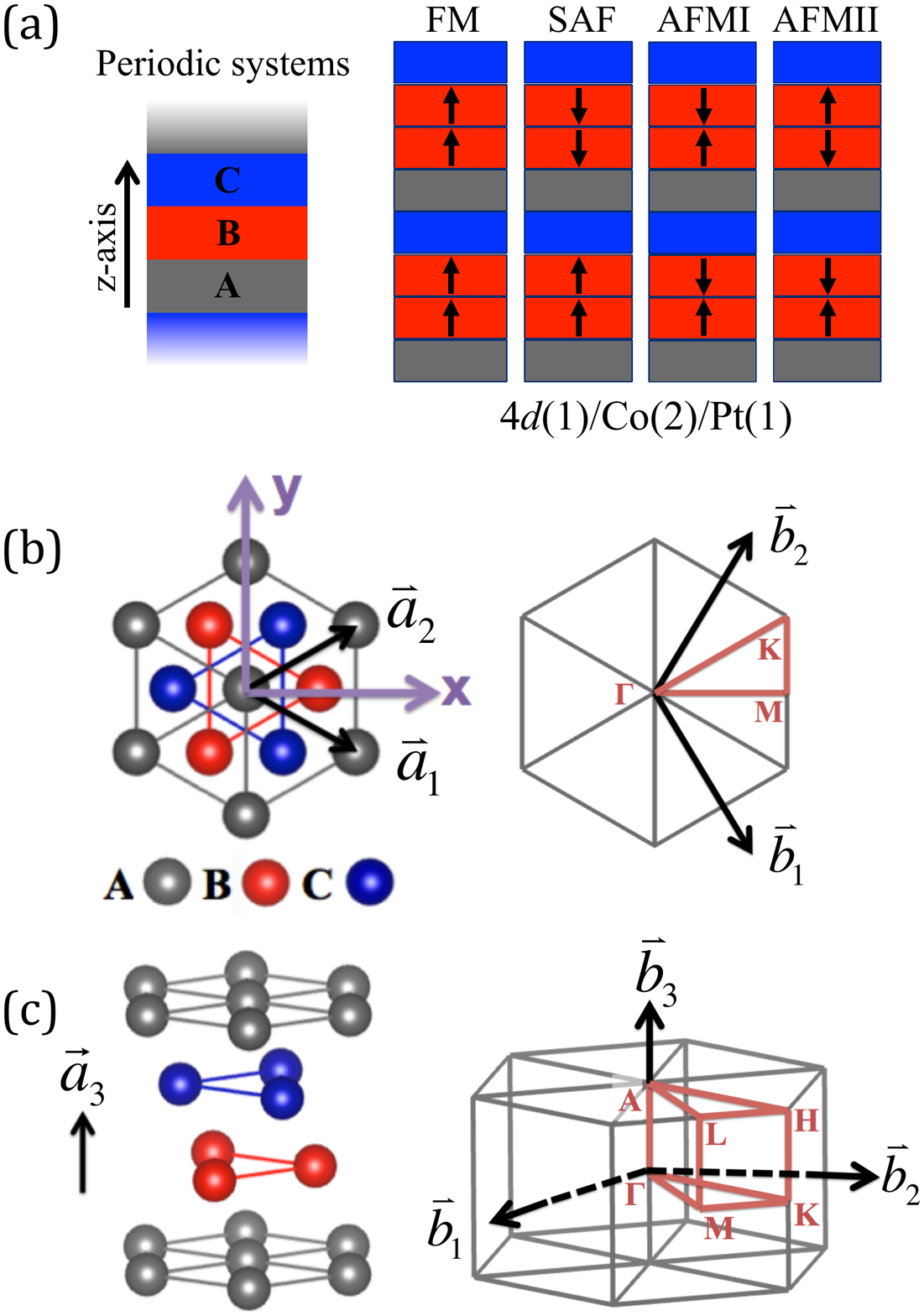}
\caption{(Color online) (a) Sketch of a periodic multilayer system made of repetitions of a trilayer structure and possible magnetic states for \{4$d$(1)/Co(2)/Pt(1)\} MMLs (FM = ferromagnet, SAF = synthetic antiferromagnet, AFMI = antiferromagnet I, AFMII = antiferromagnet II). (b) Arrangement of atoms in plane (b) and out of plane along the $z$ axis (c) in multilayer systems: the left panels show the atomic arrangement in real space, while the right hand side shows the Brillouin zone of the hexagonal lattice. $\mathbf{a}_1=\frac{1}{2}(\sqrt{3}a,-a)$ and $\mathbf{a}_2=\frac{1}{2}(\sqrt{3}a,a)$ indicate the $p(1\times1)$ unit cell of the chemical lattice, $\mathbf{b}_1$, $\mathbf{b}_2$ and $\mathbf{b}_3$ represent reciprocal lattice vectors for the chemical unit cell.}
\label{fig1}
\end{figure}

For the smallest systems studied here, \ie each metallic layer exhibits a thickness of just  1 monolayer (ML), we relax the size of the MML unit cell $c$ along the $z$-axis as well as all interlayer distances assuming an ABC (\ie fcc-like) stacking sequence (see Fig.~\ref{fig1}a). For a second set of calculations, we have each increased the thickness of one of the layers by another atomic layer, \eg \{Rh(2)/Co(1)/Pt(1)\}, where the numbers in parenthesis denote the number of atomic layers. For these systems, we additionally optimized the stacking sequence: There are three different possibilities to stack 4 layers, namely ABAB, ABAC and ABCB. The stacking sequence, which yields the lowest total energy, together with their structural details, are summarized in Table~\ref{table1}.

We obtain all three possible stacking sequences as structural ground states depending on whether a Co, Pt or Rh/Pd layer is added. Interestingly, irrespective of whether Rh or Pd is included in the MMLs, the same stacking sequence is obtained for a chosen combination of number of layers, \eg the stacking for \{Rh(1)/Co(1)/Pt(2)\} and \{Pd(1)/Co(1)/Pt(2)\} is the same, ABCB. Comparing the size of the unit cells in $z$-direction, it is clear that the Pd-based multilayers exhibit a larger lattice parameter $c$ than their Rh-based counterparts (by about 1\% per $4d$ layer) due to the larger atomic radius of Pd with the additional electron in the valence shell. In line with this trend are the interlayer distances between the $4d$ element and Co, $d_{4d-\mathrm{Co}}$, as well as those between $4d$ and Pt, $d_{\mathrm{Pt}-4d}$. However, the Co-Pt interlayer distance is significantly reduced (by 1--2\%) if Pd is included as third element as compared to Rh. This finding highlights the possibility to modify the hybridization between Co and Pt just by the presence of another element.

\subsection{Magnetic properties}

\subsubsection{Magnetic moments}\label{sec:results:mm}

The local magnetic moments of Co and the induced moments of the nonmagnetic spacer layer atoms (Rh, Pd and Pt) are listed in Table~\ref{table2}. In first approximation one finds that the Co moments of MMLs with monolayer thick Co films are very stable and about 2~$\mu_\mathrm{B}$, irrespective of the local environment determined by additional Rh, Pd, or Pt atoms and are reduced to about 1.85~$\mu_\mathrm{B}$ for MMLs with Co doublelayers. On a finer scale one finds that the Co moments are 2--5\% smaller as neighbors of Rh in comparison to Pd or Pt. The reduction of the intraatomic exchange interaction of Co due to the presence of Rh will be also discussed in the upcoming subsection \ref{subsubsection:SS} where it appears again as reduced interatomic exchange interaction in terms of the spin stiffness within the Co layer, when Rh is adjacent to Co rather than Pd or Pt. The induced moments in the non-magnetic atomic layers adjacent to Co align parallel to Co. The induced moments of Pt are slightly larger in Pd based MML as compared to Rh based MML. The Pt atoms next nearest neighbor to Co have induced moments that are already quite small with one order of magnitude smaller than the ones of Pt adjacent to Co. Rh has one electron less than Pd or Pt and thus has more holes that can be polarized and subsequently the induced Rh moments are larger than the ones of Pd and Pt. As we will see later in section \ref{subsubsection:DMI} we find that the sizes of the induced magnetic moments of the otherwise nonmagnetic spacer layer elements Rh, Pd, or Pt are totally uncorrelated with values of the DMI that are contributed by them.

\subsubsection{Interlayer exchange coupling $\&$ magnetic order in between the magnetic layers}

\begin{figure}[H]
\centering
\includegraphics[width=60mm]{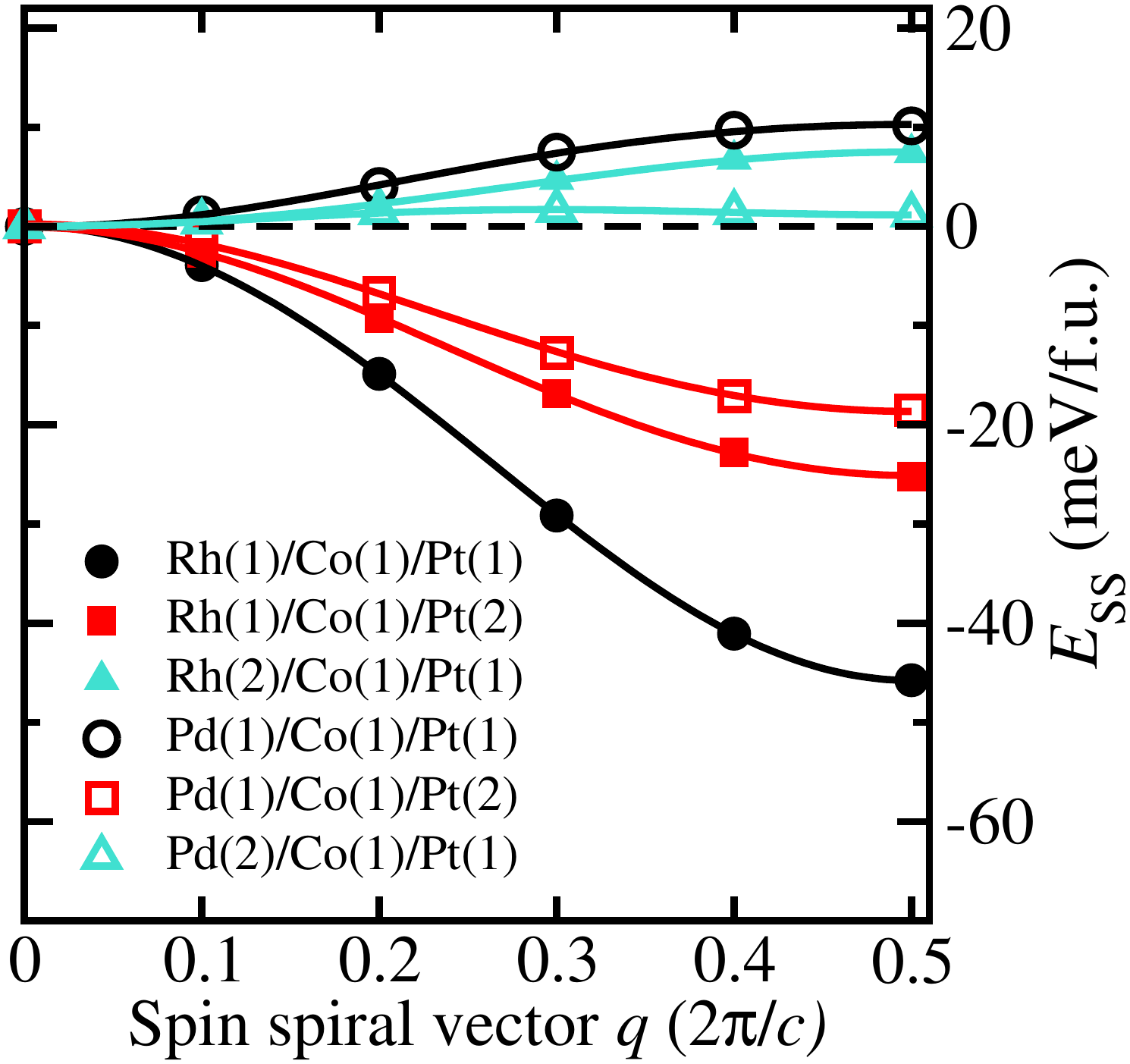}
\caption{(Color online) The energy dispersion of spin spirals with $\mathbf{q}$ along $\Gamma$-A direction to explore the interlayer exchange coupling. The least-squares fits are performed to obtain the exchange constants $J_n$.}
\label{fig2}
\end{figure}

We next investigate the interlayer exchange coupling (IEC) between the magnetic Co layers across the Rh(Pd)/Pt spacer layers. The results are summarized in Fig.~\ref{fig2} and Table~\ref{tab:iec}. Co-layers in Rh-based MMLs tend to form a synthetic antiferromagnet, whereas systems including Pd exhibit mostly a ferromagnetic IEC, with interesting exceptions: \{Rh(2)/Co(1)/Pt(1)\} exhibits a ferromagnetic IEC, and \{Pd(1)/Co(1)/Pt(2)\} an antiferromagnet IEC.

\begin{table}
 \caption{Exchange interactions parameters ($J_n$) and resulting interlayer magnetic order.  $J>0 (<0)$ denotes (anti-)ferromagnetic interaction. SAF = synthetic antiferromagnet, FM = ferromagnet. \label{tab:iec}}
 \begin{ruledtabular}
  \begin{tabular}{ccccc}
         & MML & mag.\ order&\multicolumn{1}{c}{$J_1$} & \multicolumn{1}{c}{$J_2$} \\
         & {}  & {}         &  (meV)                   &  (meV)                    \\ \hline
  \multirow{3}{*}{Rh} & \{Rh(1)/Co(1)/Pt(1)\} & SAF      &$\phantom{}-45.8$ & $\phantom{-}0.9$ \\ 
                      & \{Rh(1)/Co(1)/Pt(2)\} & SAF &$\phantom{}-25.1$ & $\phantom{}-0.5$ \\
                      & \{Rh(2)/Co(1)/Pt(1)\} & FM       &$\phantom{-} 7.6$ & $\phantom{}-0.3$ \\  \hline
  \multirow{3}{*}{Pd} & \{Pd(1)/Co(1)/Pt(1)\} & FM       &$\phantom{-}10.3$ & $\phantom{-}0.7$ \\ 
                      & \{Pd(1)/Co(1)/Pt(2)\} & SAF &$\phantom{}-18.9$ & $\phantom{}-0.6$ \\
                      & \{Pd(2)/Co(1)/Pt(1)\} & FM       &$\phantom{-} 1.1$ & $\phantom{-}1.0$ \\
  \end{tabular}
 \end{ruledtabular}
\end{table}

Expressing the energetics of the interlayer exchange coupling in terms of the Heisenberg pair interaction with coupling parameter $J_{0n}$ (see Eq.~\ref{eq:iec:ham}), we can describe the data very well with a nearest-neighbour model: The next-nearest neighbor interaction $J_2$ is at least one order of magnitude smaller than that between nearest neighbors, $J_1$ (see Table~\ref{tab:iec}). Only for the case of \{Pd(2)/Co(1)/Pt(1)\}, where the spin-spiral dispersion is very flat for the whole high-symmetry line $\overline{\Gamma \mathrm{A}}$, $J_1$ and $J_2$ are of comparable magnitude.

\begin{table}
 \caption{Energies in meV per f.u.\ relative to a reference state for several collinear configurations (see text and Fig.~\ref{fig1}a) in magnetic multilayers \{4$d$(1)/Co(2)/Pt(1)\}. \label{tab:2Co:energies}}
 \begin{ruledtabular}
  \begin{tabular}{l r r}
           & \centering $4d=\mathrm{Rh}$  & $4d=\mathrm{Pd}$ \\ \hline
     FM    &  39 \phantom{ Rh} & 0  \phantom{ Pd} \\
     SAF   &  0  \phantom{ Rh} & 16 \phantom{ Pd} \\
     AFMI  &  384\phantom{ Rh} & 524\phantom{ Pd} \\
     AFMII &  386\phantom{ Rh} & 487\phantom{ Pd}
  \end{tabular}
 \end{ruledtabular}
\end{table}

\begin{figure}[H]
\centering
\includegraphics[width=85mm]{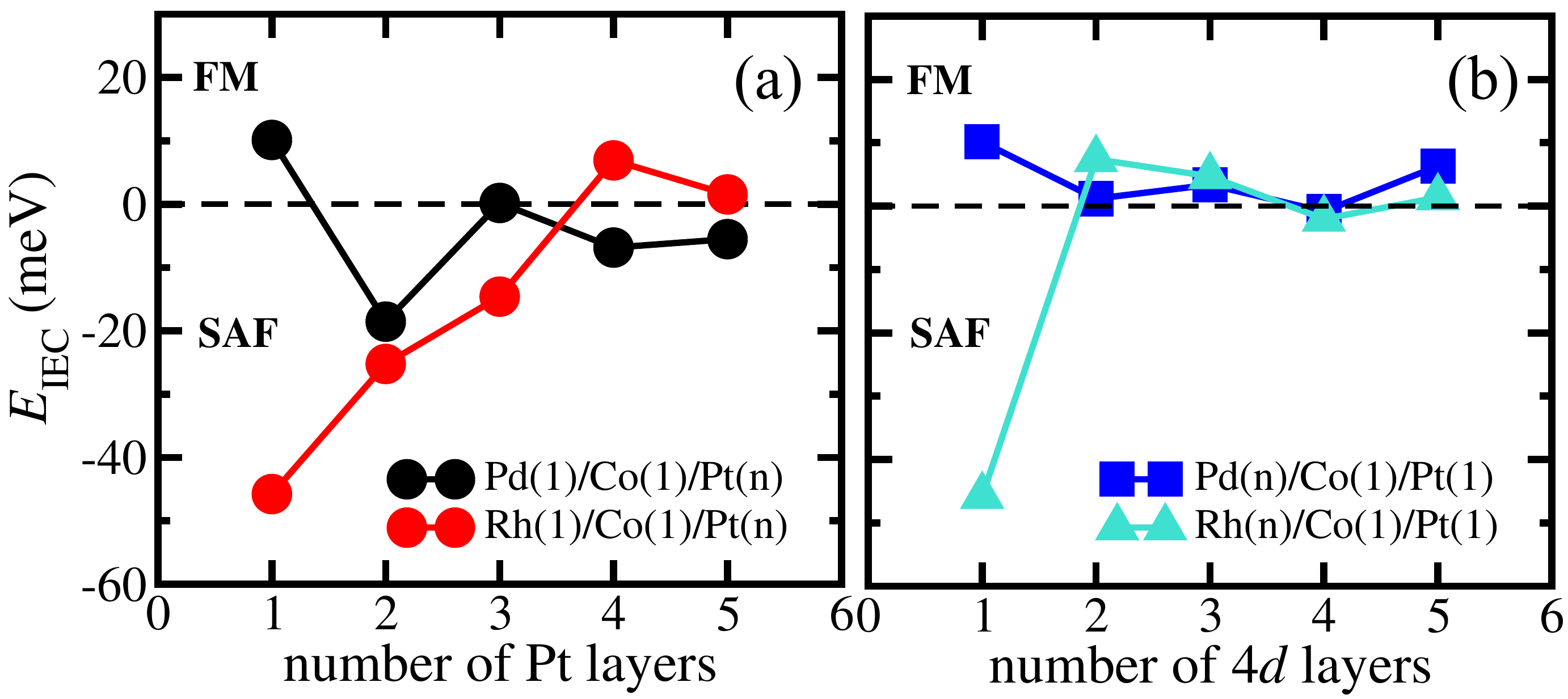}
\caption{(Color online) The interlayer exchange coupling energies $E_\mathrm{IEC}$ between the magnetic Co layers versus the number of Pt layers (a) and 4$d$ layers (b) in $\{ 4d(1)/\mathrm{Co}(1)/\mathrm{Pt}(n)\}$ MMLs.}
\label{fig3}
\end{figure}

We next explore the possibility to tune the IEC by modifying the thicknesses of the Co layers and the Rh(Pd)Pt spacer layers. At first we change the thickness of the Co layer and go from Co monolayer to the doublelayer systems, \ie \{4$d$(1)/Co(2)/Pt(1)\} and calculate the energy of 4 different collinear magnetic states, termed FM, SAF, AFMI and AFMII (see Fig.~\ref{fig1}a). As expected, the AFMI and AFMII states are several hundred meV higher in energy than FM and SAF, and unattainable by experiment, due to a large direct ferromagnetic exchange of Co (see Tab.~\ref{tab:2Co:energies}). However, the IEC is one order of magnitude smaller and the sign depends on the 4$d$ element: the lowest energy for MMLs containing Rh is a synthetic antiferromagnet, whereas Pd induces a ferromagnetic coupling. 

At second, we vary the thickness of Pt, Rh and Pd spacer layers between n=1$,\dots, 5$ atomic layers fixing the thickness of all other layers at one atomic layer. We made reasonable assumptions on the stacking sequence for these systems, but fully relaxed the interlayer distances. A typical RKKY-type oscillatory behavior is observed for \{Pd(1)/Co(1)/Pt(n)\} (see Fig.~\ref{fig3}a), with a fast oscillation period and quick decay as function of $n$. In contrast, the \{Rh(1)/Co(1)/Pt(n)\} multilayers show a much larger oscillation period and slower decay. Increasing the thicknesses of the $4d$ materials (\{$4d$(n)/Co(1)/Pt(1)\}, see Fig.~\ref{fig3}b), the MMLs prefer a ferromagnetic coupling for $n\geq2$ with energy differences lower than 10~meV.

The tendency to mediate antiferromagnetic IEC in Rh-based multilayers is similar to the effect of Ru in Co-based giant magnetoresistance (GMR) materials\cite{33Colis}. Overall, the quite complex behavior observed here is governed by the details of the electronic structure, such as the Fermi surface of involved spacer materials\cite{34Bruno}.

\subsubsection{Magnetic anisotropy energy}\label{sec:results:mca}

\begin{table*}
 \caption{The interlayer order of magnetic Co layers (FM = ferromagnetic, SAF = synthetic antiferromagnet),interlayer exchange coupling energies ($E_\mathrm{IEC}$), the magnetic moment of Co atoms ($M_\mathrm{Co}$) and induced moments of $4d$ ($M_{4d}$) and Pt ($M_\mathrm{Pt}$) atoms, the spin stiffness constant ($A$), the DMI constant ($D$), the MAE constant ($K$) and the reduced parameter $\kappa$ for several magnetic multilayers. The number in parenthesis denotes the thickness in atomic layers. $D>0 (<0)$ refers to  left(right) handed chirality. $K>0 (<0)$ refers to the out-of-plane (in-plane) easy axis.\label{table2}}
 \begin{ruledtabular}
  \begin{tabular}{cccccccccc}
Systems & \multicolumn{1}{c}{Interlayer} & \multicolumn{1}{c}{$E_\text{IEC}$} & \multicolumn{1}{c}{$M_{4d}$} & \multicolumn{1}{c}{$M_\mathrm{Co}$} & \multicolumn{1}{c}{$M_\mathrm{Pt}$} & \multicolumn{1}{c}{$A$} & \multicolumn{1}{c}{$D$} & \multicolumn{1}{c}{$K$} & $\kappa$ \\
    {}  & \multicolumn{1}{c}{order}      &     (meV)    & $\mu_\mathrm{B}$ & $\mu_\mathrm{B}$ & $\mu_\mathrm{B}$ & (meV $\mathrm{nm}^2$/{f.u.}) & (meV nm/{f.u.}) & (meV/{f.u.}) & {}  \\  \hline
\{Rh(1)/Co(1)/Pt(1)\}    & SAF & $-45.8\phantom{-}$ & 0.41 & 1.92  &  0.17  & 136 & $-1.62\phantom{-}$  & 0.79 & \phantom{3}66   \\
\{Rh(1)/Co(1)/Pt(2)\} &SAF  & $-25.2\phantom{-}$ &  0.55   & 1.96 &  0.31/-0.01   & 147 & 2.12  & 0.54 & \phantom{3}29     \\
 \{Rh(1)/Co(2)/Pt(1)\} & SAF & $-39.4\phantom{-}$ &  0.34 & 1.84/1.86 & 0.21  & 214 & 0.69  & 0.66 & 481     \\
 \{Rh(2)/Co(1)/Pt(1)\} & FM  & $\phantom{9}7.4$ & $-0.12$/0.28 &1.89  &  0.24  & 102  & $-1.38\phantom{-}$ & $-0.38\phantom{-}$ & \phantom{3}33      \\ \hline
\{Pd(1)/Co(1)/Pt(1)\}    & FM  & 10.1 & 0.33 &2.01  &  0.34  & 126 & 4.39  & 1.51 & \phantom{3}16      \\
{\{Pd(1)/Co(1)/Pt(2)\}} & SAF & $-18.5\phantom{-}$ & 0.29 & 2.02 &  0.33/0.03   & 167 & 7.36  & 2.01 & \phantom{3}10       \\
\{Pd(1)/Co(2)/Pt(1)\} & FM  & 15.7 & 0.29 & 1.88/1.88 & 0.29 & 266 & 5.42  & 0.37 & 5.4       \\
\{Pd(2)/Co(1)/Pt(1)\} & FM  & $\phantom{5}1.2$ & 0.30/0.34 &2.05 &  0.38   & 162 & 5.11  & 2.06 & \phantom{3}21      \\ 
  \end{tabular}
 \end{ruledtabular}
\end{table*}

The total magnetic anisotropy energy (MAE) coefficients $K$, which comprises the contributions from spin-orbit coupling and classical magnetic dipole-dipole interactions, are presented in Tab.~\ref{table2}.

The MAE in these (111)-oriented MMLs is uniaxial, $E_\mathrm{MAE} = -K (\mathbf{m} \cdot \hat{e}_z)^2$. We calculate the MAE as energy difference between states with magnetization pointing in plane (along the $x$ axis) and out-of-plane (along $z$).

As presented in Tab.~\ref{table2}, nearly all MMLs have an out-of-plane easy axis, which is typical for materials comprising Co and Pt \cite{38Nakajima}. The MAE of Pd-based MMLs are generally very large, since the Pd(111) and Co/Pt(111) interfaces show a strong perpendicular magnetic anisotropy (PMA). Adding one more Pd or Pt layer to the thinnest MML stack considered here increases the PMA by about 35\%, whereas adding a Co-layer leads to a considerable reduction of PMA. Generally, the MAE in Rh-based MMLs is smaller by a factor 2--3, and can even turn the easy-axis in plane (see \{Rh(2)/Co(1)/Pt(1)\} in Tab.~\ref{table2}). Based on Eq.~\ref{eq:3}, we can see that the low magnitude of magnetic anisotropy coefficients will facilitate the emergence of cycloidal spirals.

\subsubsection{Spin stiffness}
\label{subsubsection:SS}

To extract the spin stiffness in these multilayers, we calculate the energy dispersion of homogeneous spin spirals with spin-spiral vector $\mathbf{q}=\mathbf{q}_0 + \mathbf{q}_\mathrm{eff}$, where $\mathbf{q}_0$ represents the lowest-energy state as determined by the IEC (\ie $\mathbf{q}_0=\Gamma$ for FM and $q_0=\mathrm{A}$ for SAF, see Tab.~\ref{tab:iec} and Tab.~\ref{tab:2Co:energies}). For $q_\mathrm{eff}$ we chose a vector that lies in the plane of the MML, and points towards the $\Gamma$-M direction.\cite{footnote} Only $q_\mathrm{eff}$ determines the non-collinear order within a layer, and we obtain as period length of a spin-spiral $\lambda = 2\pi/q_\mathrm{eff}$.

\begin{figure}[H]
\centering
    \includegraphics[width=85mm]{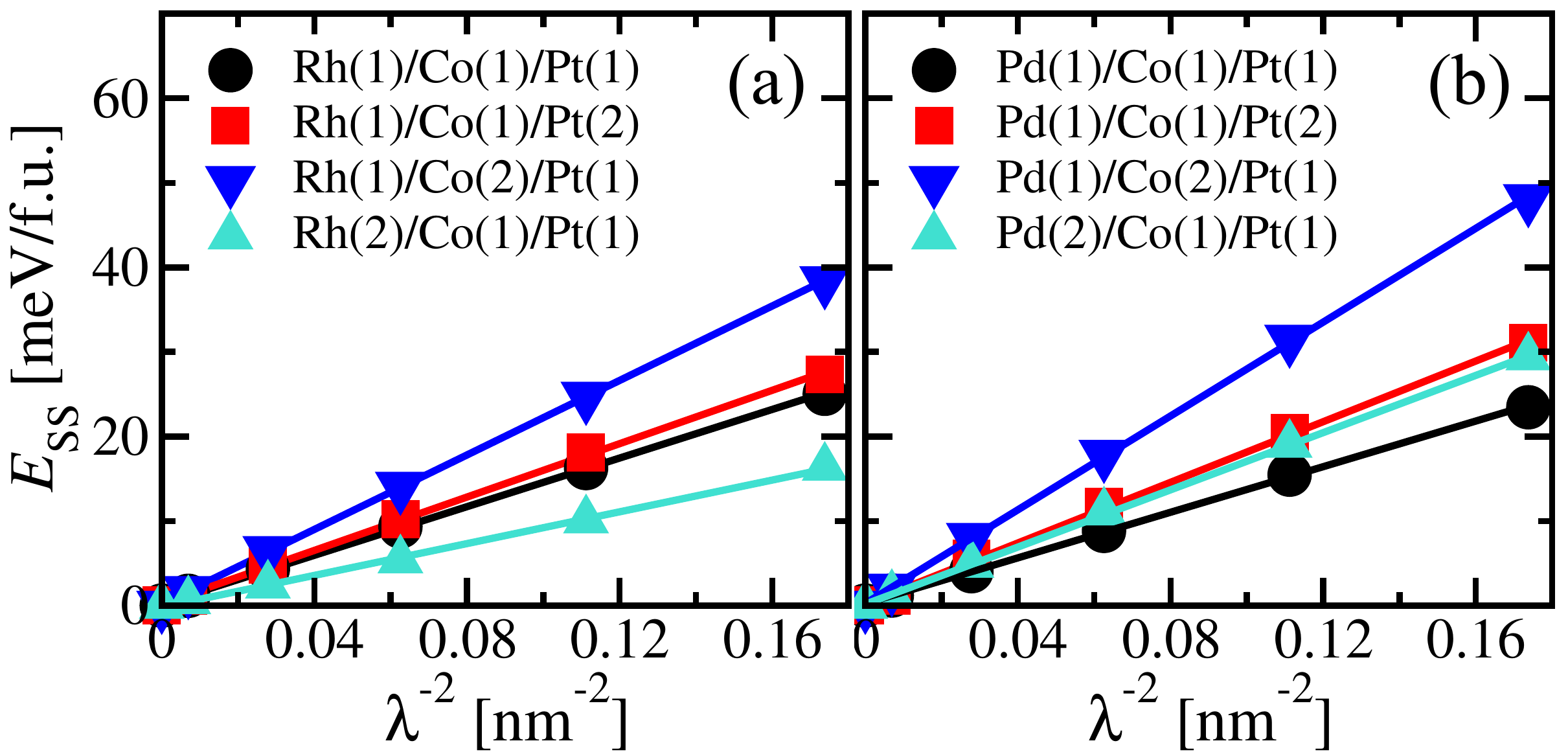}
\caption{(Color online) The spin-spiral energy ($E_\mathrm{ss}$) dispersion is shown as a function of $\lambda^{-2}$ ($\lambda = 2\pi\left| q \right|^{-1}$ is the wavelength of the spin spiral) for (a) the Rh/Co/Pt and (b) the Pd/Co/Pt MMLs. The linear fits are used to obtain the spin-stiffness $A$.}
\label{fig4}
\end{figure}

Fig.~\ref{fig4} displays the spin-spiral energy $E_\mathrm{SS}$ as function of $\lambda^{-2}$ and the spin-stiffnesses $A$ obtained as slopes (see Eq.~\ref{eq:2}) from corresponding fits are summarized in Table~\ref{table2}. It is noted that the spin stiffness in Pd/Co/Pt MMLs is mostly larger (up to 37\%) than the one in corresponding Rh/Co/Pt MMLs, and hence having in mind the formation of non-collinear magnetization textures such as skyrmions, in Pd/Co/Pt MMLs it is normally needed to overcome a larger isotropic exchange-interaction energy. This softening of the spin stiffness on Co introducing Rh to the MML is similar to the effect of Rh in Fe-based multilayers \cite{21Dupe}. The spin stiffness $A$ in the MMLs with two Co-atoms per f.u., is about twice as large, which simply stems from the fact that the spin-stiffness scales with the number of Co layers and thus with the amount of the magnetic volume in the multilayer.

\subsubsection{Dzyaloshinskii-Moriya Interaction}
\label{subsubsection:DMI}

\begin{figure}[H]
\centering
\includegraphics[width=85mm]{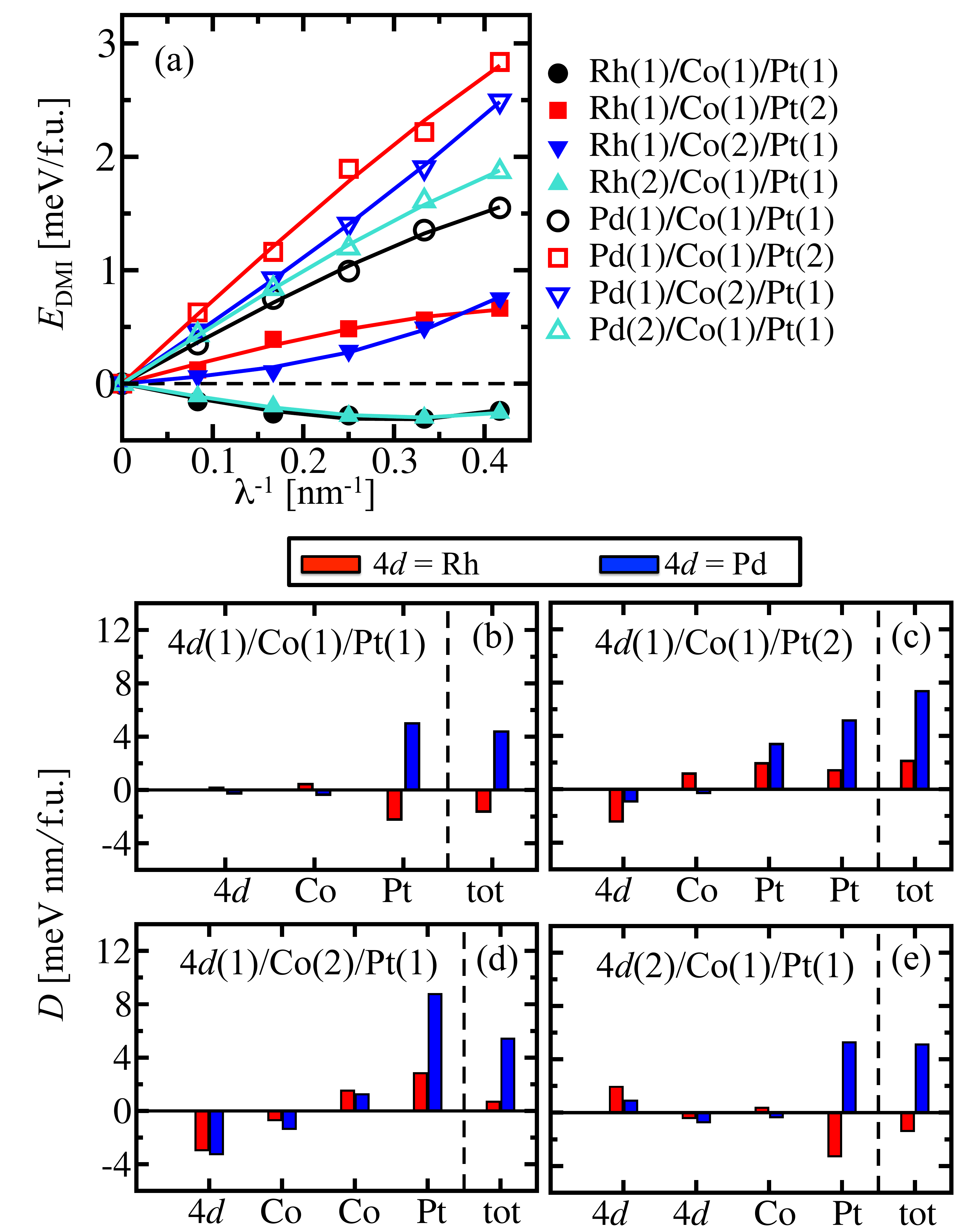}
\caption{(Color online) (a) The DMI energies ($E_\mathrm{DMI}$) of spin spirals are shown as a function of $\lambda^{-1}$ and fits to obtain the parameter $D$. (b-e) The total DMI (tot) and the contributions of different atomic layers to the DMI.}
\label{fig5}
\end{figure}

In order to extract the Dzyaloshinskii-Moriya interaction (DMI) parameter, $D$, we calculated the SOC-induced energy shift of cycloidal spin-spirals with wavevectors $\mathbf{q}_{\mathrm{eff}}$ as used in the previous section \ref{subsubsection:SS} in the vicinity of the collinear state of lowest energy. The resulting energies are collected in Fig.~\ref{fig5}a as function of $\lambda^{-1}$. We then extract the micromagnetic DMI-constants $D$ as slopes to a cubic fit of the data (see Table~\ref{table2}). According to our sign convention, $D > 0$ ($D < 0$) implies a lowering of spin-spiral energies with left-rotational (right-rotational) sense.

The Pd/Co/Pt MMLs exhibit a strong DMI with values ranging between 4.4 and 7.4~meV\,nm/f.u., which are of similar strength and of the same chirality as a single Co/Pt interface ($7.1$~meV\,nm/f.u., taken from Freimuth \etal\cite{43Freimuth} and accounted for a factor $1/(2\pi)$ due to different definitions). Interestingly, the DMI in the synthetic antiferromagnet \{Pd(1)/Co(1)/Pt(2)\} is the largest of the here investigated systems. This promotes the idea of obtaining stable, small skyrmions in a SAF. 

In comparison to the Pd-based systems, the DMI in Rh/Co/Pt MMLs is much weaker. In addition, there is a strong variation of magnitude of the DMI as function of the individual layer thicknesses. Even the sign of the DMI can change and can become negative so that magnetic structures  with  right-rotational sense is preferred.

This seems surprising, since the main contribution to the DMI is often attributed to the Co/Pt interface, which is always present in the MMLs under consideration. In order to obtain a deeper insight, we computed the layer-resolved contributions to the DMI by activating the SOC atom by atom. As evident from the Fig.~\ref{fig5}(b-e), we indeed see that the largest contributions stem from Pt atoms. However, we find them to be very sensitive with respect to the number of atomic layers as well as the chemical element (Rh or Pd) that interfaces with Pt and Co. As an example, the Pt-induced DMI is as large as +9~meV\,nm/f.u.\ in Pd(1)/Co(2)/Pt(1), but only about +3~meV\,nm/f.u.\ in the same stack with Pd replaced by Rh, and for Rh(2)/Co(1)/Pt(1), the Pt-contribution to the DMI even changed sign ($-3$~meV\,nm/f.u.). The system Rh(1)/Co(1)/Pt(2) illustrates that the modification of the DMI values of Pt is really a nonlocal effect, as the DMI contribution of both Pt atoms are effected, although only one Pt atom interfaces to Rh. Hence, the overall puzzling behavior of the total DMI originates from the Pt atoms.

Also surprising is the fact that in Pd(1)/Co(1)/Pt(2), the second Pt layer with little spin polarization (see subsection~\ref{sec:results:mm}), the one adjacent to the Pd metal, yields the largest contribution to the DMI, even larger than the Pt layer interfacing Co directly. This is different to ultra-thin Co films on Pt(111)\cite{8Yang}, where the DMI is originating nearly exclusively from the atomic layer at the Co/Pt interface. Similarly to Yang \etal \cite{8Yang}, we do not find any direct correlation between the DMI and the size of induced magnetism of Pt. The induced magnetic moment of Pt stems from a direct hybridization with Co atoms, enhanced by the intra-atomic susceptibility of Pt. Exemplary for Pd(1)/Co(1)/Pt(1), we investigated the origin of the large magnetic moment of Pt amounting to $0.34~\mu_\mathrm{B}$ by switching off the intra-atomic exchange enhancement of Pt at each step of the self-consistent cycle of the DFT calculations by suppressing the magnetic part of the exchange correlation potential, $B_\mathrm{xc}=0$. We find that the induced magnetic moment reduces to $0.24~\mu_\mathrm{B}$, which is interpreted as the magnetic moment of Pt that results from a spin-polarized hybridization of the Pt wave functions with Co, or in other words the polarization of Pt by the nonlocal and non-enhanced susceptibility of Pt. Within this model, the total DMI and the contribution of the Pt layer are 4.61 and 5.12~meV\,nm/f.u., respectively, \ie almost unchanged in comparison to those including the intra-atomic exchange enhancement (4.39 and 4.99~meV\,nm/f.u., respectively). Our findings are consistent with the results of Sandratskii \cite{Sandratskii}, where the local moment of Pt was totally suppressed by an external constraint. It can be concluded that the asymmetry of the Pt wave function is the origin of the DMI but not the induced local magnetic moment.

A trend that we observe is, that the Pt-contributions in Pd-based MMLs are larger than the ones in Rh-based MMLs, which might be attributed to the smaller interlayer distance  between Co and Pt atoms (see discussion in Sec.~\ref{sec:results:structure}), facilitating a stronger hybridization and DMI. An additional factor is the charge transfer and the respective potential gradient that impacts the size of the DMI. Considering the CoPtPd trilayer as part of the Pd based MML and taking into account that Pt and Pd are isoelectronic, the charge and potential gradients are clearly at the CoPt interface. This is different for the CoPtRh trilayer in Rh based MMLs. Co and Rh are isoelectronic and from the viewpoint of charge transfer, Pt is positioned in an electronically  much more symmetric environment and smaller DMI is expected.  Furthermore, we observe that the contributions from the $4d$-layers are sizable, but of different sign to the ones from Pt, and hence decrease the total DMI by up to 50\% in Rh(1)/Co(2)/Pt(1).

\subsubsection{Discussion: Magnetic in-plane order}

Based on the spin-stiffness $A$, DMI constant $D$, and the magnetic anisotropy coefficient $K$ determined from the \abinitio\ calculations, we deduce the effective parameter $\kappa$ (see Eq.~\ref{eq:3}) to determine the magnetic ground state within the magnetic layers. The results are listed in Table~\ref{table2} and the fact that  $\kappa > 1$ for all systems reveals a collinear magnetic order for all \{$4d$/Co/Pt\} MMLs considered here. 

However, for Pd(1)/Co(2)/Pt(1) the effective parameter $\kappa$ ($\kappa=5.4$) is relatively close to the transition towards a spin-spiral state and even closer to the metastability of skyrmions. Indeed, for a material with a similar $\kappa$ (an Fe double layer on W(110) with $\kappa=4.8$\cite{9Heide}), cycloidal N\'{e}el-type walls induced by external magnetic fields have been observed experimentally\cite{40Bergmann} and the appearance of meta-stable two-dimensional chiral magnetic solitons, in this case an anti-skyrmion, has been predicted\cite{41Hoffmann}. Therefore, we conjecture that a Pd(1)/Co(2)/Pt(1) magnetic multilayer is a promising candidate for spintronic applications. Recently, Pollard \etal reported the interesting result that chiral spin structures including skyrmions have been observed in Co/Pd multilayers experimentally at room temperature\cite{42Pollard}, which further supports our conclusion.

\section{Conclusion}
We investigated by means of density functional theory calculations the structural and magnetic properties of (111) oriented $4d$/Co/Pt magnetic multilayers. We focused on properties like interlayer exchange coupling, magnetic anisotropy, spin stiffness and Dzyaloshinskii-Moriya interaction, all relevant for the investigation of one- and two-dimensional (meta-)stable chiral magnetic solitons. We targeted $4d$ transition-metal modified Co/Pt multilayers with the aim to tune the exchange interaction independently of the spin-orbit related properties of the Co/Pt interface. We selected Rh and Pd as $4d$ elements as Rh (Pd) is isoelectronic to Co (Pt). We studied multilayers with one and two atomic layers of Co and varied the different chemical components of the spacer layer between 1 and 5 atomic layers.
 
The number of atomic planes of the individual magnetic or non-magnetic layers influences the stacking sequence. For example the Co double-layer induces an hexagonal stacking of the MML. As function of the thickness of the spacer layers we find ferromagnetic and synthetic antiferromagnetic interlayer coupling except for Pd($n$)/Co(1)/Pt(1), where only a ferromagnetic coupling was found for all Pd thicknesses investigated.

All investigated combinations show an out-of-plane easy axis, with the exception of Rh(2)/Co(1)/Pt(1), which exhibits an easy-plane anisotropy.
As a general trend, Pd-based systems exhibit a slightly larger spin-stiffness and a much larger Dzyaloshinskii-Moriya interaction as compared to Rh-based MMLs. In combination with a reduced magnetic anisotropy energy resulting in a low $\kappa$, we conclude that the Pd(1)/Co(2)/Pt(1) magnetic multilayer is a promising candidate for spintronic applications, in which the meta-stable skyrmions can be expected at the presence of an external magnetic field, which is consistent with recent experimental results of Pollard \etal \cite{42Pollard}.

While Pd alters very little the DMI at the Co/Pt interface, Rh has a strong non-local effect, modifying the Co/Pt DMI even if Rh is not a direct neighbor of Pt interfacing Co. Here the idea of modifying the exchange interaction and the DMI independently by introducing interfaces of Co with Pt and Co with a $4d$ metal breaks down. This analysis will motivate further investigations of chiral properties of Co/Pt based magnetic multilayers and provides guidance for multiscale explorations and experimental search for skyrmions in these systems.

\section{Acknowledgments}

We thank M. Hoffmann, F. Lux and G. Bihlmayer for fruitful discussions. We gratefully acknowledge financial support from the MAGicSky Horizon 2020 European Research FET Open project (\#665095) and the DARPA TEE program through grant MIPR (\# HR0011831554) from DOI, as well as computing resources at the supercomputers JURECA at Juelich Supercomputing Centre and JARA-HPC from RWTH Aachen University.


\begin{thebibliography}{99}
\bibitem{footnote}{Actually, all in-plane directions are equivalent in the micromagnetic limit due to the high C$_\mathrm{3v}$ symmetry of the lattice.}
\end{thebibliography}
\end{document}